# The Case for an Industrial Policy Approach to AI Sector of Pakistan for Growth and Autonomy

**Atif Hussain**

*Institute of Quality & Technology Management; and Center for Industrial Policy & Market Analysis, University of the Punjab, Pakistan*
atif.iqtm@pu.edu.pk

**Rana Rizwan**

*Solutyics (Pvt) Limited*
rana.rizwan@solutyics.com

## ABSTRACT

This paper argues for the strategic treatment of artificial intelligence (AI) as a key industry within Pakistan's broader industrial policy framework, underscoring the importance of aligning it with national goals such as economic resilience and preservation of autonomy. The paper starts with defining industrial policy as a set of targeted government interventions to shape specific sectors for strategic outcomes and argues for its application to AI in Pakistan due to its huge potential, the risks of unregulated adoption, and prevailing market inefficiencies. The paper conceptualizes AI as a layered ecosystem, comprising foundational infrastructure, core computing, development platforms, and service/product layers, supported by education, government policy, and R&D. The analysis highlights that Pakistan's AI sector is predominantly service-oriented, with limited product innovation and dependence on foreign technologies, posing risks to economic independence, national security, and employment. To address these challenges, the paper recommends educational reforms, support for local AI product development, initiatives for indigenous cloud and hardware capabilities, and public-private collaborations on foundational models. Additionally, it advocates for public procurement policies and infrastructure incentives to foster local solutions and reduce reliance on foreign providers. This strategy aims to position Pakistan as a competitive, autonomous player in the global AI ecosystem.

## 1. INTRODUCTION

In the 21st century, artificial intelligence (AI) is rapidly becoming a key driver of economic growth, technological innovation, and societal transformation (Qin et al., 2024; Trabelsi, 2024). Countries around the world are recognizing AI's potential to revolutionize industries, enhance productivity, and create new economic opportunities. For developing nations like Pakistan, the adoption of AI holds the promise of significant productivity and efficiency gains as well as economic diversification opportunities (Chinasa, 2023; Fan & Qiang, 2024; Mishra et al., 2023). However, the AI revolution also poses grave risks of not only widening the gap between developed and developing economies but also creating unequal gains within developing countries (Alonso et al., 2020; Korinek & Stiglitz, 2021; Mannuru et al., 2023; UN SDG:Learn, n.d.).

In Pakistan, AI applications, particularly in areas such as "precision agriculture," are frequently highlighted as solutions to long-standing productivity challenges (Danial, 2024). However, if adopting these technologies means relying on imports from other countries, it may negatively impact Pakistan's balance of payments. Moreover, The reliance on foreign technologies can exacerbate economic imbalances, deepen technological dependency, and limit domestic innovation in developing countries (Fabayo, 1996) whereas reducing technological dependence on foreign countries can increase total factor productivity (Zhang et al., 2023). Over-reliance on foreign technologies may also pose threats to national security in the emerging techno-geopolitical context. Thus, without a strategic approach, AI could become a trap that undermines Pakistan's economic and technological sovereignty and development.

Given these challenges, this essay argues that AI should be treated as a strategic industry in Pakistan, warranting the activation of state's agency. Rather than merely a tool for improving efficiency, AI should be integrated into a broader industrial policy framework. By treating AI as a strategic sector, Pakistan can align its AI strategy with national economic objectives—such as boosting exports, attracting remittances and foreign direct investment (FDI), creating jobs, and enhancing economic resilience. This approach would mitigate the risks of technological dependency while enabling Pakistan to leverage AI for long-term economic growth, technological independence, and sustainable development.

The essay begins by defining industrial policy, then conceptualizes AI as an industry, discusses the reasons why Pakistan should adopt an industrial policy approach for AI, and outlines the key features of a potential industrial policy for AI in Pakistan.

## 2. WHAT IS INDUSTRIAL POLICY?

The concept of industrial policy lacks a universally agreed-upon definition, with many interpretations emerging from different scholars (Aiginger & Rodrik, 2020; Ambroziak, 2017;

Warwick, 2013). Broadly speaking, industrial policy refers to a comprehensive set of government interventions aimed at strategically shaping and enhancing the domestic economy by targeting specific industries, firms, or sectors. These efforts involve deploying various instruments and measures designed to improve the structural performance, competitiveness, and long-term growth of the business sector, with the overarching goal of achieving specific economic or strategic objectives (Criscuolo et al., 2022; McDonald et al., 2024). Such interventions, which can utilize both demand- and supply-side instruments (Criscuolo et al., 2022), may be applied economy-wide or focused on particular sectors, commonly referred to as horizontal and vertical policies, respectively (Maloney & Nayyar, 2018). Historically, industrial policy has focused on manufacturing, but that notion is now being challenged with arguments in favor of using the same approach for service (Juhász et al., 2024).

Historically, liberal economic thought has been critical of industrial policy, emphasizing the primacy of free trade and market mechanisms. However, the limitations of unregulated markets have become increasingly evident, and recent examples of successful interventions have reignited interest in industrial policy (Chang & Andreoni, 2020; Vaughan, 2017). Even critics acknowledge that, at least in part, industrial policy has contributed to economic growth (Cheang, 2024). In fact, the data shows that developed countries (generally more liberal) are more heavy users of industrial policy measures (Evenett et al., 2024).

In recent years, especially with the rapid advancement of technologies like artificial intelligence (AI), countries are recognizing the need for such policies to maintain economic autonomy and competitiveness (Criscuolo et al., 2022; Fontana & Vannuccini, 2024; Mariotti, 2024). Most recent data shows that geopolitical concerns are one of the main motives of industrial policy initiatives (Evenett et al., 2024). As global powers vie for technological leadership and dominance (Konrad, 2024), other regions are prompted to pursue industrial policy to safeguard their technological sovereignty, reduce dependency on dominant powers, and preserve autonomy (Edler et al., 2023; Zhao, 2024).

## 3. A CONCEPTUALIZATION OF AI AS AN INDUSTRY

Making AI any target of policy action would require some conceptualization of its components, processes, and boundaries. Historically, the term has been associated chiefly with manufacturing, however, now service sectors are also understood to be and referred to as industries (e.g., entertainment industry and education industry).

Industry has generally been defined as a distinct group of businesses with sufficiently similar processes, technical resources and skill requirements, producing close substitutes (*Definition of INDUSTRY*, 2024; Nightingale, 1978). However, some authors argue that defining an industry solely by substitutable products is inadequate, particularly in the information age, and instead recommend emphasizing the concept of "activity networks"(Munir & Phillips, 2002). Moreover, as Bouwens et al. (2018) note, 'industry' is not an entity but an operational concept that helps with understanding and should be defined according to the aim of the analysis.

Therefore, in this paper, we conceptualize AI Industry as a layered arrangement of actors and entities as depicted in Figure 1 below. Each layer builds upon the one below it, providing distinct resources, technologies, and services that together support the development, deployment, and commercialization of AI technologies. Below, we provide a detailed explanation of each layer in the AI industry, illustrating how they collectively contribute to the industry as a whole.

| | | Government | | | | |
|---|---|---|---|---|---|---|
| Layer4 | Service & Product | Services | | | Products | |
| Layer3 | Platforms & Models | Foundation Models | | Specialized Platforms | | Education, Training, R&D |
| Layer2 | Computing Core | Specialized Hardware | Cloud Platforms | Development Languages, Frameworks & Tools | | |
| Layer1 | Digital Backbone | Internet | General Purpose Hardware | Network Technologies | Payment Platforms | |

*Figure 1: AI Industry*

### 3.1. First Layer (Digital Backbone)

Layer 1 represents the digital infrastructure that serves as the foundation for the entire AI ecosystem. This includes the internet, general-purpose hardware, network technologies, and payment platforms. The internet plays a crucial role in enabling data flow, connectivity, and access to digital resources on a global scale. General-purpose hardware refers to computing devices, such as servers and computers, that perform essential computational tasks for various applications. Network technologies further support this by ensuring stable and efficient communication, which is critical for deploying AI systems in real-time and distributed environments. Payment platforms provide the financial infrastructure necessary for the commercialization of AI, facilitating the transactions and exchanges required for the operation of AI services. Together, these components create the foundational infrastructure that supports the computational and connectivity requirements of the layers above.

### 3.2. Second Layer (Computing Core)

Layer 2 encompasses the core computing technologies and resources necessary to develop AI applications. Specialized hardware, such as GPUs (Graphics Processing Units) and TPUs (Tensor Processing Units), forms an essential part of this layer by providing the computational power required for complex AI/ML tasks. Cloud platforms are a key enabler in this layer, offering on-demand, scalable computational resources that organizations can use without investing in expensive on-premise hardware. These platforms allow AI models to be developed and tested in scalable environments, lowering barriers to entry for smaller players in the industry. Furthermore, this layer includes development languages, frameworks, and tools that are pivotal for AI development. Programming languages like Python, along with machine learning frameworks such as TensorFlow and PyTorch, empower developers to design, train, and experiment with AI models efficiently. Layer 2, therefore, provides the computational environment that forms the technical backbone for developing AI solutions.

### 3.3. Third Layer (Platforms and Models)

Layer 3 comprises platforms and models that are essential for the development, deployment, and scalability of AI technologies. Foundation models, such as GPT-4 or BERT, are pre-trained, large-scale models that can be adapted for specific tasks across different domains. These models have reduced the time and cost required to develop new AI solutions by providing a reusable base for various applications. Specialized platforms are also a fundamental part of this layer. These platforms enable the development and deployment of AI technologies by providing integrated environments that encompass all the necessary tools and processes for building, deploying, and managing AI models. Specialized platforms are crucial for facilitating the end-to-end lifecycle of AI technologies, from training and tuning models to deploying them at scale. These specialized platforms ensure that AI technologies can be deployed efficiently, consistently, and securely, bridging the gap between development and operationalization.

### 3.4. Fourth Layer (Services and Products)

Layer 4 includes the services and products that emerge from previous layers, bringing AI solutions to market. This layer involves AI/ML development, analytics services, and all related aspects of development such as front-end, back-end, and mobile deployment. AI consultancy and training services as well as cybersecurity services are also included in this layer. Data annotation services which play an essential role in preparing high-quality labeled datasets, crucial for model training, are also a part of this layer. Layer 4 also includes AI products, ranging from chatbots and recommendation engines to sophisticated enterprise AI systems that drive decision-making and automation. These services and products are the culmination of all prior efforts, translating foundational technologies and platforms into real-world applications that meet user needs and drive value.

### 3.5. Overarching Components: Government, Education, Training, and R&D

Overarching all four layers are key components that ensure the AI ecosystem grows responsibly and sustainably. Government plays a regulatory role, defining policies that guide AI development, funding foundational research, providing oversight to ensure ethical and fair AI use, facilitating the industry through various industrial policy interventions, setting priorities according to national goals, and ensuring national security. The government's role is essential in building a regulatory framework that keeps pace with the rapid evolution of AI technologies. Education, training, and R&D are also essential components that span all layers of the AI ecosystem. Education and training ensure the workforce is adequately prepared to work with AI technologies, while ongoing R&D pushes the boundaries of what AI can achieve, continuously innovating across the entire stack. These overarching elements create a supportive environment that enables technological growth, mitigates ethical concerns, and provides the expertise needed to leverage AI effectively across all sectors of the economy.

## 4. THE STATE OF AI INDUSTRY IN PAKISTAN

The AI industry in Pakistan is gradually evolving, with different companies participating across the various layers of the AI ecosystem. To understand the state of the AI industry in Pakistan, it is helpful to assess the layers and components in which Pakistani companies are currently operating.

Generally, most Pakistani companies involved in AI are active in Layer 4, focusing on providing AI services rather than product development. This includes AI/ML development services, analytics solutions, and consultancy, especially for international clients. Many companies in Pakistan are engaged in developing custom AI models, offering data annotation services, and providing software development services that integrate AI into existing systems. However, there is limited emphasis on creating proprietary AI products, with most companies prioritizing service-based offerings over product innovation. While operating in Layer 4 is beneficial, for greater economic gains, companies should also place more focus on product development, which can attract initial investments as well as ensure continued future cash flows. Developing proprietary products will also allow Pakistani firms to establish a more competitive position globally. These companies often help clients in industries such as healthcare, finance, and retail to adopt AI-driven solutions for automation and decision-making support.

In addition to focusing on services, Pakistani companies should aim for greater involvement in the lower layers of the AI ecosystem, such as Layers 1, 2, and 3. From the perspective of maintaining autonomy and protecting national security, it is important for Pakistani companies to play more prominent roles and gain ownership in these foundational layers. Greater involvement in Layers 1, 2, and 3 will help ensure technological sovereignty, resilience, and national security, reducing dependency on external infrastructure and services.

Currently, in Layer 3, Pakistani companies are contributing primarily by providing workforce and supporting international efforts, rather than developing their own specialized AI platforms. These companies often utilize cloud-based services, typically provided by international providers, to facilitate the training and deployment of AI models. Although foundational model development is limited, some firms are leveraging existing foundation models and customizing them for specific use cases within Pakistan. For sustainable growth, developing indigenous specialized platforms could bolster national expertise and reduce reliance on foreign technologies.

In Layer 2, Pakistani companies are largely dependent on third-party cloud providers and specialized hardware infrastructure from global suppliers, as domestic production of such hardware is minimal. Companies leverage international cloud platforms to meet their computational needs, which enables AI model development but at the cost of reliance on external entities. The use of development languages, frameworks, and tools such as TensorFlow, PyTorch, and Scikit-learn is widespread, though these are typically sourced from the global open-source community. Strengthening local cloud and hardware capabilities could significantly enhance Pakistan's AI independence and infrastructure.

Layer 1 infrastructure in Pakistan still presents significant challenges, particularly in terms of internet quality and access to reliable hardware. Network technologies and payment platforms are gradually improving, but substantial development is still required to fully support the needs of a scalable AI ecosystem. Limited access to high-performance computing hardware and inconsistent connectivity continue to hinder the growth of AI infrastructure. Strengthening Layer 1 infrastructure will be crucial in creating a robust foundation that allows growth across all other layers of the AI stack.

To summarize, the state of AI in Pakistan is characterized by an active presence in service-oriented roles, especially in Layer 4, with limited activity in product development and foundational layers. For greater economic gains and technological independence, Pakistani companies should prioritize developing proprietary AI products and increase their involvement in the foundational layers of the ecosystem. Doing so will not only create more sustainable cash flows but also foster national resilience, technological sovereignty, and competitive global positioning.

## 5. WHY AN INDUSTRIAL POLICY APPROACH FOR AI IN PAKISTAN?

The adoption of an industrial policy approach to artificial intelligence (AI) in Pakistan is not merely a policy option but a strategic imperative. AI holds vast potential for driving economic growth, advancing societal development, and ensuring national security. However, without coordinated state intervention, the associated risks and market inefficiencies could undermine these opportunities. To fully capitalize on AI's benefits and mitigate its risks, Pakistan must adopt a strategic industrial policy approach, addressing two key considerations: the significant potential of AI and the substantial risks of unregulated adoption to both the economy and autonomy of the country. Each of these is described below and summarize in Table 1 and 2.

### 5.1. The Potential is Great

AI presents enormous opportunities for Pakistan's economy, workforce, and global competitiveness. With a strategic industrial policy, Pakistan can harness AI to drive sectoral transformation, stimulate economic diversification, and significantly boost growth.

#### *5.1.1. Export of AI and Related Services*

As global demand for AI-driven solutions continues to grow, Pakistan is well-positioned to tap into this expanding market. The country has already demonstrated notable progress in its IT and IT-enabled services sector, with exports crossing the $1 billion mark in 2018 and rapidly increasing to $2.5 billion in 2022. This figure is projected to reach approximately $3.2 billion in 2024 (KHAN, 2024; SBP, 2023). These gains have largely been the result of proactive government measures, and a similar approach could be applied to AI exports. AI's rise has created additional demand for cloud computing, front- and back-end development, and mobile applications. Furthermore, policy interventions could help diversify Pakistan's export destinations and expand its market share in existing regions, where it currently holds a small presence.

### 5.1.2. Efficiency Gains and Competitiveness

AI is often heralded as a solution to improving productivity and performance in key sectors such as agriculture, healthcare, manufacturing, and education in developing countries like Pakistan. The integration of AI into these sectors can optimize resource use, enhance decision-making, and streamline operations (Aderibigbe et al., 2023; Chinasa, 2023, 2023; Fan & Qiang, 2024; Kshetri, 2020). For example, AI-driven precision agriculture technologies utilize algorithms to analyze data from soil sensors, drones, and satellite imagery, providing farmers with real-time recommendations on irrigation, fertilization, and pest control (Siddiqui et al., 2023). This can significantly improve crop yields, reduce waste, and lower operational costs, contributing to greater food security and sustainable resource use in Pakistan, where agriculture is a key contributor to GDP and employment. Additionally, AI adoption in small and medium enterprises (SMEs) could enhance efficiency and productivity (Hussain & Rizwan, 2024).

### 5.1.3. Attracting Foreign Direct Investment (FDI)

Countries that position themselves as AI innovation hubs have the potential to attract substantial foreign direct investment (FDI), as multinational corporations seek to invest in emerging AI markets. Big tech companies are expanding their presence in developing countries (Chinasa, 2023), although Pakistan has yet to fully benefit from this trend. A proactive industrial policy could attract such investment. Moreover, AI startups are gaining traction globally, and Pakistan has seen some success in this area(Nadeem, 2024). However, the country ranks only 71st in the Global Startup Index as of 2024 (StartupBlink, 2024), leaving significant room for improvement. Government actions, such as those seen in India, which attracts about 5% of global greenfield FDI inflows in ICT (Arthur D Little, 2021), could help Pakistan achieve similar success.

### 5.1.4. *Increasing Remittances by Exporting Skilled AI Talent*

Remittances from Pakistan's overseas workforce currently account for approximately 10% of the country's GDP (De Padua et al., 2024). While there are concerns about the exodus of skilled labor, exporting AI talent can yield long-term benefits through increased remittances. In addition to the direct economic contributions, skilled expatriates can also foster increased trade links and business collaborations. Countries such as Nigeria have implemented talent export programs to maximize such benefits (*NATEP*, n.d.).

### 5.1.5. *Job Creation and Skill Development*

The growth of the AI sector in Pakistan would generate significant employment opportunities. This includes technical roles such as AI developers, data scientists, and cloud computing specialists, as well as non-technical jobs in areas such as product management and data annotation (Hussain, 2024). By promoting AI skill development through targeted educational programs and training initiatives, Pakistan can create a skilled workforce capable of driving future innovation and economic growth.

| Table 1: Summary of Opportunities ||
|---|---|
| **Opportunities** | **Description** |
| Export of AI and Related Services | Pakistan can increase its exports by tapping into the growing global demand for AI-driven solutions, building on its strong IT services exports. Targeted policies could help diversify export markets and increase Pakistan's market share in emerging regions. |
| Efficiency Gains and Competitiveness | AI can enhance productivity in critical sectors like agriculture, healthcare, manufacturing, and education, optimizing resource use, reducing waste, and increasing crop yields. This transformation can make Pakistani industries more competitive globally and improve food security. |
| Attracting Foreign Direct Investment (FDI) | Positioning Pakistan as an AI hub could attract foreign investments as multinational companies increasingly invest in AI markets. Government policies similar to those in other developing countries could enhance Pakistan's ranking in the global startup ecosystem and attract more FDI in ICT. |
| Increasing Remittances by Exporting Skilled AI Talent | By exporting skilled AI professionals, Pakistan can boost remittances, contributing to GDP and fostering international trade links. Implementing targeted talent export programs could maximize these benefits and create business collaborations abroad. |
| Job Creation and Skill Development | AI sector growth would create jobs in technical roles (AI developers, data scientists) and non-technical roles (product managers, data annotators). Promoting AI-focused educational programs can build a skilled workforce to support economic growth and innovation. |

## 5.2. The Risks are Great

Despite AI's potential, its unregulated adoption poses significant risks that could exacerbate Pakistan's existing vulnerabilities and create new challenges for economic and national security. Without strategic oversight, these risks could outweigh the potential benefits.

### *5.2.1. Technological Dependence on Foreign Countries*

Technological dependence on foreign suppliers has far-reaching implications beyond mere economic concerns. The control exerted by suppliers of digital technologies can undermine a country's sovereignty, allowing foreign powers to extract resources and shape the economic, cultural, and political landscape (Kwet, 2019). Furthermore, the dominance of big tech firms can make the economy reliant on their platforms, exacerbating technological dependency ((Yeşilbağ, 2022). Technological dominance is increasingly being weaponized for geopolitical leverage, eroding the autonomy of dependent states and creating vulnerabilities in the global power structure (Choer Moraes & Wigell, 2022; Ringhof & Torreblanca, 2022). Without political action, Pakistan risks losing its technological sovereignty.

### *5.2.2. Rising Import Bill for AI Technologies*

Relying on imported AI technologies brings explicit costs (such as initial purchase) and implicit costs (such as ongoing maintenance) that can strain Pakistan's balance of payments. The over-reliance on foreign technologies, without developing local alternatives, could significantly increase the import bill of a country (Amaechi, 2020; Erber, 1983), further exacerbating its trade deficit.

### *5.2.3. Job Displacement and Economic Inequality*

AI-driven automation poses a real threat to employment in sectors traditionally dependent on human labor. For example, AI-powered robots in manufacturing may reduce the comparative advantage that developing countries like Pakistan enjoy through low-cost labor. As a result, investments may shift towards robotics in developed countries, where manufacturing can be reshored, leaving developing countries with closed factories and rising unemployment (Alonso et al., 2020; Lu & Zhou, 2021). Even within Pakistan, AI automation may disproportionately benefit capital over labor, leading to increased economic inequality as jobs are lost to machines (Lu & Zhou, 2021). Emerging evidence suggests that these technological advancements may lead to reshoring operations in developed economies, diverting investments away from developing regions (Bilbao-Ubillos et al., 2024; Kinkel et al., 2023; Pillich, 2024). Government intervention is crucial to mitigate these risks and ensure a balanced transition.

### 5.2.4. *National Security Vulnerabilities*

While AI technologies have applications that could bolster national security (Al-Suqri & Gillani, 2022; Mallick, 2024; Yu & Carroll, 2021), reliance on foreign providers poses significant risks. These external actors may engage in mass surveillance or data theft, compromising sensitive government or security information (Noor, 2020; Retter et al., 2020). Additionally, AI models developed by foreign companies may inadvertently (or intentionally) undermine national security by acting against the interests of the dependent state. A strong policy framework is needed to develop domestic AI capabilities, safeguarding national security.

| Table 2: Summary of Risk ||
|---|---|
| **Risks** | **Description** |
| Technological Dependence on Foreign Countries | Relying on foreign suppliers for AI technologies can compromise Pakistan's sovereignty and expose it to geopolitical manipulation. Dominance by foreign tech firms could increase dependency, allowing foreign powers to exert influence over Pakistan's economy and politics, risking technological sovereignty. |
| Rising Import Bill for AI Technologies | Dependence on imported AI technologies increases costs and maintenance expenses, straining Pakistan's balance of payments. Without local alternatives, the growing import bill for AI technologies could worsen the trade deficit. |
| Job Displacement and Economic Inequality | AI-driven automation could displace jobs in labor-intensive sectors, potentially leading to job losses and increased economic inequality. As manufacturing automates, countries with a low-cost labor advantage, like Pakistan, may lose out to robotics, leading to unemployment and reshoring of operations to developed nations. |
| National Security Vulnerabilities | Reliance on foreign AI providers poses risks of surveillance and data theft, compromising sensitive national data. AI models developed abroad could contain biases that may undermine Pakistan's national interests. A policy framework is necessary to build local AI capabilities, protecting national security interests. |

## 6. POSSIBLE POLICY INTERVENTIONS

Various policy interventions can be made to achieve the goals and mitigate the risks discussed previously. These interventions are discussed in the following sections and summarized in Table 3.

## 6.1. Education as a Foundational Layer

Education serves as the bedrock for enabling every aspect of the AI ecosystem. The government should take substantial measures to reform curricula across various disciplines—not just within information technology—to align education with practical applications in the real world. This necessitates:

### 6.1.1. *Updating Curricula*

Incorporating mandatory AI and data science components not only for STEM fields but also integrating interdisciplinary applications in sectors such as agriculture, business, and healthcare. Such initiatives should emphasize practice-oriented learning. Project-based instruction, internships, and industry partnerships should be embedded into these programs to ensure students acquire practical, hands-on experience. Universities should be encouraged to collaborate with local industries to develop courses that specifically address the workforce requirements and existing skill gaps across sectors.

### 6.1.2. *Skills Programs and Scholarships:*

The development of specialized skills training programs and the establishment of scholarship opportunities are imperative to cultivating talent in AI and data science. Particular attention should be given to providing financial support for underprivileged students pursuing AI-related education. Existing initiatives, such as the Prime Minister's Youth Skill Development Program, should be expanded to encompass more advanced and specialized AI training. Additionally, new scholarships focusing on AI and technology disciplines ought to be introduced.

The challenge of employability remains significant, with reports indicating that only approximately 10% of IT graduates in Pakistan are considered employable (P@SHA, 2022). Collaborative initiatives with industry leaders should aim to develop targeted boot camps and short courses to facilitate the rapid upskilling of individuals for roles within the emerging AI economy. Existing programs like the Higher Education Commission (HEC) Digital Learning Skills Development Program, which are currently limited to students, should be broadened to include the wider populace. With only 14% of IT companies in Pakistan reportedly having formal learning partners, the government should also consider launching initiatives to address these deficiencies, fostering partnerships that enhance industry-specific learning.

Furthermore, the government's existing internship programs could be expanded to encourage international companies to recruit interns from Pakistan. Such initiatives would not only provide global exposure to local talent but also ensure that Pakistani graduates are adequately equipped with skills relevant to international markets.

### 6.1.3. University Funding

Allocating dedicated funding to universities for AI-related research is essential to enable educational institutions to function as innovation hubs for AI development and application. The establishment of competitive research grants, alongside the promotion of centers of excellence in AI, will enhance the capacity of universities to contribute to the nation's AI landscape. Initiatives such as the National Center for Artificial Intelligence (NCAI) should be tasked with the development of localized solutions to national challenges. Moreover, expanding the network of research and development centers focusing on AI will help foster a culture of innovation, ultimately strengthening the national AI ecosystem.

## 6.2. Layer 4: Services and Products
### 6.2.1. Services

The government already facilitates all IT service companies, including AI service providers, with tax incentives and support for attending international events, additional measures should be implemented. However, a critical issue to address is the disparity in tax treatment between freelancers and companies, as the latter face a broader range of taxes while freelancers often pay a significantly lower rate. Addressing this inequity is crucial for fostering a balanced competitive environment.

Moreover, in addition to facilitating companies in attending international events, to build credibility, the Pakistan Software Export Board could introduce a certification program for AI service providers, enhancing market confidence. Similarly, it can play its role in promoting partnerships between Pakistani firms and international AI service providers which will further boost technical expertise and create opportunities for the exchange of best practices.

Similarly, Government-backed AI service export councils can play an essential role by providing strategic guidance, negotiating trade agreements, and assisting domestic companies in accessing new markets. Moreover, integrating AI services into broader export promotion strategies will create a cohesive and synergistic approach to international market development.

Finally, AI services should also be incorporated into various government loan programs, ensuring these companies have sufficient access to capital for growth.

### 6.2.2. Products

Encouraging product development by local companies should be a major goal of the policy interventions. Products for the local market will reduce dependence on imported products whereas products for the global market will ensure cash inflows. Several steps can be taken. For example, targeted product development grants should be introduced, prioritizing products that have a greater local impact or global appeal. Initiatives like "ignite" are already in place and should receive more funding to broaden their scope.

Industry-academia collaboration is crucial for transforming academic research into market-ready products. This can be facilitated by establishing dedicated offices within universities to streamline commercialization processes. Additionally, AI product incubators should be set up to provide early-stage startups with mentorship, seed funding, and access to prototyping facilities, enabling them to bring their innovations to market more effectively.

To ensure global visibility, participation in international AI product expos should be actively supported, providing Pakistani companies with platforms to showcase their developments, attract investment, and form strategic partnerships. Creating an AI Product Development Award will also celebrate and recognize successful innovations, fostering a culture of excellence, competition, and motivation.

To improve access to funding and attract both local and international investment, an AI-focused venture capital fund should be established to provide seed and growth capital to promising startups. Expanding microfinance programs will further help support small businesses and entrepreneurs adopting or developing AI solutions, promoting financial inclusion across the sector. Additionally, a government-backed loan guarantee scheme should be introduced to mitigate financial risks for lenders, thus encouraging more financial institutions to invest in AI-related projects. Finally, the Pakistan Startup Fund, which has been established to provide more confidence to foreign investors, can offer additional incentives for investing in AI startups, helping attract more local and international funding.

### 6.3. Layer 3: Platforms and Foundation Models
#### *6.3.1. Creation of Foundation Models:*

It is imperative to support the development of foundation models tailored to Pakistan's specific needs through localized R&D efforts or collaborations with international technology giants. Dedicated funding should be allocated for research initiatives in critical sectors like agriculture, healthcare, finance, and natural language processing, including language models for Urdu and regional languages. Establishing AI research centers in collaboration with local universities, in addition to the already existing initiatives like National Centre for Artificial Intelligence (NCAI), can promote research excellence in foundational model development. Public-private partnerships should also be fostered, where local companies can contribute to the development and training of these models using indigenous datasets. To facilitate this, secure data-sharing frameworks must be put in place, allowing high-quality data to be accessed by researchers while ensuring adherence to data privacy regulations. Moreover, a national data repository should be established, providing researchers and startups access to diverse datasets necessary for building accurate AI models.

#### *6.3.2. Collaboration with International Tech Giants:*

Establishing strategic partnerships with global technology firms to co-develop and customize foundation models that cater to local needs is essential. These collaborations should include knowledge transfer programs, ensuring that the skills and expertise acquired through these partnerships are disseminated across local academic and research institutions. Negotiations should focus on agreements that include provisions for training Pakistani engineers and researchers, thereby building local capacity in foundational AI technologies. Additionally, international technology giants should be incentivized to establish regional R&D centers in Pakistan, serving as hubs of innovation for the development of localized AI models. Providing regulatory support and tax incentives to these companies will encourage long-term investments in local R&D activities.

### 6.4. Layer 2: Computing Core
#### *6.4.1. Local Cloud Facilities:*

Local cloud infrastructure is crucial for national autonomy and security, and several steps can be taken to achieve these goals. The National Centre for Big Data and Cloud Computing should expand its operations to provide subsidized access to computing power for startups, universities, and researchers. This expansion should include specialized nodes for high-performance computing (HPC) and AI workloads to support advanced research needs. Collaboration with educational institutions should also be strengthened by offering cloud credits to support faculty- and student-led research projects, thereby bridging the gap between academia and industry-grade computing.

Encouraging the use of local cloud resources for research will gradually strengthen the domestic cloud infrastructure, develop skills, and maintain the confidentiality and control of projects critical to Pakistan. Additionally, the government should introduce regulations to require that sensitive data in certain sectors be stored in local data centers. This demand-side intervention will promote the localization of cloud computing, reduce reliance on foreign services, and accelerate the expansion of national cloud capabilities.

### 6.4.2. *Local Hardware Manufacturing:*

Incentivizing local manufacturing or assembly of AI-specific hardware is crucial for reducing reliance on imports and developing a local ecosystem. A "Make in Pakistan" campaign should be introduced to support electronic hardware manufacturing for AI, offering subsidies, reduced tariffs, and training programs for local manufacturers. Some manufacturers, such as Haier Pakistan already have a manufacturing presence in Pakistan. They should be encouraged to open new assembly lines for AI-specific hardware, thus providing training and job opportunities for local talent. Dedicated industrial zones for electronic hardware manufacturing should also be established, complete with streamlined regulatory processes and tailored incentives for companies engaged in AI hardware production. This initiative would meet local hardware demands, create export opportunities, and help modernize Pakistan's industrial base, currently dominated by low-tech sectors like textiles.

### 6.5. Layer 1: Digital Backbone
### 6.5.1. *Broadband Expansion through PPPs*

Expanding broadband access is fundamental for the growth of AI technologies. This can be achieved through public-private partnerships, targeting underserved areas, and offering subsidies or tax incentives to telecom companies. Ambitious targets should be set for 5G rollout in urban centers, and fiber-optic connectivity should be extended to rural regions to support data-intensive AI applications. A monitoring body should be established to ensure the timely completion of broadband infrastructure projects, with transparent performance reports to uphold accountability.

### 6.5.2. *Tax Incentives for Infrastructure Development:*

Waiving import duties on essential infrastructure equipment for digital connectivity and offering tax holidays for companies investing in rural internet infrastructure will be instrumental in enhancing digital access. Grants should be provided to telecom providers that meet specific milestones in enhancing digital coverage, ensuring even the most remote areas have reliable internet access. Additionally, targeted grants should be introduced to support satellite internet projects that bring connectivity to regions where terrestrial solutions are less feasible, such as remote or mountainous areas.

## 6.6. Integration with Other Industries and Encouragement of Local Procurement
### 6.6.1. *Local Procurement Policy:*

Introducing a mandatory policy whereby a certain percentage of AI technologies used in critical industries—such as healthcare and agriculture—must be sourced locally will encourage participation from domestic companies in national projects, thereby promoting technology sovereignty. A "Buy Pakistani AI" initiative should be launched to prioritize locally developed AI solutions in government contracts, providing local firms with the incentive to innovate and scale.

### 6.6.2. *AI Adoption in Key Industries:*

Industry-specific AI adoption strategies must be developed, particularly for sectors like agriculture, where precision agriculture technologies can enhance productivity, or healthcare, where AI-based diagnostics can lead to better patient outcomes. Targeted grants and tax incentives should be made available to companies integrating AI to improve their operational efficiency. Establishing AI transformation funds for traditional industries would allow the government to co-invest alongside private firms, creating co-investment opportunities for AI adoption projects.

Training programs should be developed specifically for industry professionals to foster a deeper understanding of AI integration into existing workflows, thereby bridging the gap between traditional industries and cutting-edge AI technologies. AI innovation labs should be set up and made accessible to industry professionals, offering hands-on workshops and opportunities for collaboration to drive AI-driven improvements across sectors. By adopting AI, these industries will not only make gains in productivity and efficiency but will also generate a demand pull that will lead to development of AI industry.

### *6.6.3. Public Procurement*

Public procurement can be a strategic industrial policy tool to accelerate the growth of Pakistan's AI sector. By implementing a "Local AI First" policy, the government can create a stable demand for domestic AI solutions, providing local companies with the opportunity to build credentials through government projects. These credentials can enhance their competitiveness in international markets, as they can demonstrate proven success in large-scale, public-sector deployments.

Furthermore, to create a fairer playing field for private AI firms, the government should reconsider the role of provincial information technology boards, which currently develop software and digital solutions directly for government use. This practice often excludes private companies from competing in key projects. Shifting the focus of these boards to a regulatory role—rather than a development one—would encourage private sector participation, stimulate innovation, and prevent crowding out. This approach would ensure that government projects leverage private sector expertise, helping the local AI industry mature and compete globally.

## 7. CONCLUSION

A strategic industrial policy approach is essential for Pakistan to fully harness the transformative potential of artificial intelligence while securing economic sovereignty, enhancing competitiveness, and fostering sustainable growth. Treating AI as a strategic sector can help Pakistan mitigate risks such as technological dependence, job displacement, and national security vulnerabilities. The proposed framework advocates for interventions across foundational infrastructure, core computing, development platforms, and AI services, supported by a robust education system, effective government policy, and continuous research initiatives. Emphasizing local product development and strengthening foundational technology capabilities will reduce reliance on foreign technologies, build economic resilience, and secure a competitive edge in the global AI landscape. Through education reforms, public-private partnerships, incentives for local procurement, and investment in indigenous infrastructure, Pakistan can establish an AI ecosystem that not only addresses domestic needs but also positions the country as a significant player in international AI markets. This strategic, multi-layered approach to AI development enables Pakistan to leverage AI as a catalyst for economic transformation, job creation, and technological sovereignty, paving the way for a resilient, self-sufficient, and innovative future.

| Table 3: Summary of Proposed Policy Measures ||
|---|---|
| **Area** | **Measures** |
| Education, Skills, R&D | Update curricula to include AI and data science across disciplines.<br>Expand skills programs and scholarships for AI-related education.<br>Broaden HEC programs to the general populace.<br>Expand internship opportunities, including international placements.<br>Increase funding for AI-related university research.<br>Promote centers of excellence and R&D centers in |
| Service & Product Layer | Address tax disparities between freelancers and companies.<br>Introduce certification programs for AI providers.<br>Establish government-backed AI export councils.<br>Incorporate AI services into government loan programs.<br>Introduce grants for AI product development with local impact.<br>Enhance industry-academia collaboration for commercialization.<br>Set up AI incubators for startups.<br>Support participation in international AI expos.<br>Establish an AI Product Development Award.<br>Create an AI-focused venture capital fund and expand microfinance.<br>Provide incentives through the Pakistan Startup Fund for AI investment. |
| Platforms and Foundation Models | Fund R&D for localized foundation models.<br>Develop secure data-sharing frameworks and a national data repository.<br>Establish strategic partnerships with global tech firms.<br>Provide incentives for global firms to establish R&D centers in Pakistan. |
| Computing Core | Expand local cloud infrastructure<br>Make local cloud available for startups and researchers.<br>Mandate local data storage for sensitive sectors.<br>Support local manufacturing and assembly of hardware.<br>Launch a "Make in Pakistan" campaign for hardware. |
| Digital Backbone | Expand broadband access through public-private partnerships.<br>Waive import duties on digital infrastructure equipment.<br>Offer tax holidays for rural internet investments.<br>Provide grants for achieving digital infrastructure milestones. |
| Integration with other industries | Introduce a local procurement policy prioritizing domestic AI solutions.<br>Offer grants and incentives for AI adoption in key industries.<br>Establish AI transformation funds to support AI integration.<br>Set up industry-specific AI training programs and innovation labs. |
| Public Procurement | Implement a "Local AI First" policy for government projects.<br>Shift provincial IT boards to a regulatory role to enhance private sector participation in government contracts. |